\documentclass[10pt,twocolumn,letterpaper]{article}
\usepackage{subcaption}
\usepackage{iccv}
\usepackage{times}
\usepackage{epsfig}
\usepackage{graphicx}
\usepackage{amsmath}
\usepackage{amssymb}

\usepackage[breaklinks=true,bookmarks=false]{hyperref}

\iccvfinalcopy 

\def\iccvPaperID{****} 

\ificcvfinal\fi

\begin{document}

\title{Studying the Impact of Augmentations on Medical Confidence Calibration}

\author{Adrit Rao$^{1, 2}$\quad Joon-Young Lee$^{3}$\quad  Oliver Aalami$^{1}$\\
{\footnotesize \tt adritrao@stanford.edu \quad jolee@adobe.com \quad aalami@stanford.edu} \\
		 $^{1}$Stanford University \qquad
             $^{2}$Palo Alto High School \qquad
		 $^{3}$Adobe Research  \\
	}


\maketitle
\ificcvfinal\thispagestyle{empty}\fi

\begin{abstract}
   The clinical explainability of convolutional neural networks (CNN) heavily relies on the joint interpretation of a model's predicted diagnostic label and associated confidence. A highly certain or uncertain model can significantly impact clinical decision-making. Thus, ensuring that confidence estimates reflect the true correctness likelihood for a prediction is essential. CNNs are often poorly calibrated and prone to overconfidence leading to improper measures of uncertainty. This creates the need for confidence calibration. However, accuracy and performance-based evaluations of CNNs are commonly used as the sole benchmark for medical tasks. Taking into consideration the risks associated with miscalibration is of high importance. In recent years, modern augmentation techniques, which cut, mix, and combine images, have been introduced. Such augmentations have benefited CNNs through regularization, robustness to adversarial samples, and calibration. Standard augmentations based on image scaling, rotating, and zooming, are widely leveraged in the medical domain to combat the scarcity of data. In this paper, we evaluate the effects of three modern augmentation techniques, CutMix, MixUp, and CutOut on the calibration and performance of CNNs for medical tasks. CutMix improved calibration the most while CutOut often lowered the level of calibration.
\end{abstract}

\section{Introduction}

\begin{figure}[h]
    \centering
    \includegraphics[width=\linewidth]{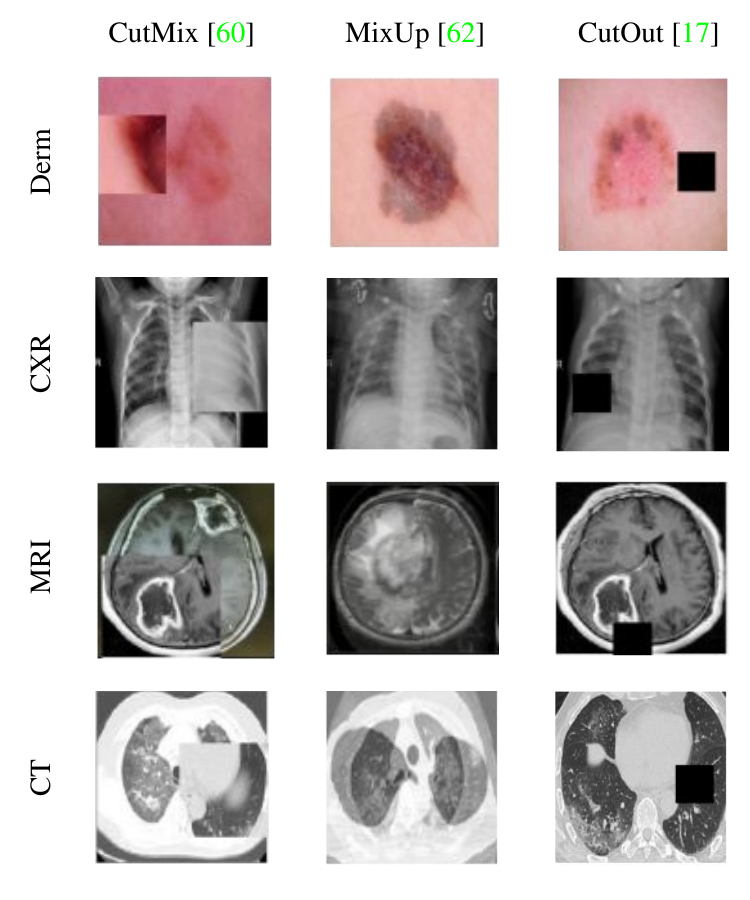}
    \caption{\textbf{Modern image augmentations applied to various medical modalities to create unseen samples.} In this study, we evaluate the effects of modern augmentations on the confidence calibration and performance of CNNs for medical image analysis. Augmentations studied are: CutMix \cite{yun2019cutmix}, MixUp \cite{zhang2017mixup}, and CutOut \cite{devries2017improved}. CNNs are trained and benchmarked on augmented datasets of the Skin, CT, CXR, and MRI medical modalities.}
    \label{fig:augmentations}
\end{figure}

The applications of computer vision to medical image analysis have been widely studied \cite{shen2017deep, litjens2017survey, lee2017deep, esteva2021deep}. In medical image analysis, models are trained to assist clinicians in the triaging of image diagnosis through the interpretation of medical images (e.g. CT, MRI, CXR, skin, etc.). The goal of these models is to increase diagnostic accuracy and efficiency by reducing manual time-consuming tasks \cite{chan2020deep}. In recent years, such models have become increasingly \textit{accurate} \cite{rajpurkar2017chexnet,park2019deep,huang2020penet,rajpurkar2020appendixnet,varma2019automated} leading to the wider adoption of computer vision-based tools in clinical settings. Currently, the \textit{accuracy} of medical image analysis models are evaluated primarily based on performance-based statistical metrics \cite{anwar2018medical}. Upon obtaining high \textit{accuracy}, these models are typically deployed in the clinical setting for validation or comparison against trained clinicians \cite{rajpurkar2018deep,bien2018deep,bellemo2019artificial,qin2019using}.

In addition to maintaining a high accuracy, it is important to ensure the reliability of these models to support safe clinical decision-making. In most medical image analysis tasks, a predicted diagnostic label and associated confidence probability are presented jointly to the clinician. A clinician must interpret this information taking into consideration not only the provided diagnosis but the certainty or uncertainty of the model to gain trust \cite{cosmides1996humans}. While \textit{accuracy} solely encompasses the ability to correctly predict a certain class, a model must provide a \textit{reliable} confidence estimation to the clinician. When using these models for diagnostic aid, confidence estimates can provide major insight to clinicians and significantly influence diagnosis. For example, a model which outputs a confidence of 98\% should provide the clinician with a higher certainty for that diagnostic label. On the other hand, a low confidence of 60\% should help the clinician consider the uncertainty and potentially revise the diagnosis. Ensuring that confidence estimates are reflective of the \textit{true correctness likelihood} of diagnostic labels is essential, creating the need for \textbf{confidence calibration} \cite{guo2017calibration}.

It has been shown that state-of-the-art convolutional neural networks (CNN) are often poorly calibrated and tend to be overconfident or \textit{overly certain} in predictions \cite{guo2017calibration}. This poses a significant problem and potentially a high risk in the medical domain. Calibrated confidence estimates are not captured in performance-based metrics commonly used to evaluate medical image analysis models \cite{anwar2018medical}. A low level of calibration can easily go unnoticed leading to improper measures of uncertainty and potentially false interpretations when deployed in the clinical setting. In the general computer vision community, various techniques have been proposed to improve the calibration, reliability, and transparency of neural networks \cite{denker1990transforming, mackay1992practical,gal2016dropout,lakshminarayanan2017simple}. Modifying certain features (e.g. network width, height, depth) of a CNN can also significantly effect confidence calibration \cite{guo2017calibration}. However, such calibration techniques are not commonly leveraged for medical image analysis as many applications perform transfer learning of pretrained CNNs \cite{kora2021transfer,ravishankar2016understanding,raghu2019transfusion}. These applications may yield high performance but could potentially lack high levels of calibration.

In recent years, various \textit{modern image augmentation} techniques have been introduced. Three notable techniques which are studied in this paper are MixUp \cite{zhang2017mixup}, CutMix \cite{yun2019cutmix}, and CutOut \cite{devries2017improved}. As implied in the names, these augmentations perform unique operations such as cutouts, blends, and mixes on images to generate unseen samples. Each of these techniques have not only benefited CNNs in terms of performance but also have significant benefits for confidence calibration. Various studies have shown the advantages of these augmentations for calibration of CNNs using standard calibration-based metrics on general computer vision benchmarks \cite{thulasidasan2019mixup, zhang2022and, carratino2020mixup, chun2020empirical, das2020empirical}. Additionally, similar to standard image augmentations (e.g. crop, rotate, flip) these techniques improve the regularization and robustness of CNNs to out-of-distribution (OOD) samples. 

Standard image augmentations \cite{krizhevsky2017imagenet, szegedy2017inception, huang2017densely} such as random scaling, rotation, and zooming of image samples are widely leveraged for medical image analysis \cite{chlap2021review}. The reason augmentations are used in the medical domain is to: increase the \textit{size} and the \textit{diversity} of datasets to improve robustness and reduce overfitting (i.e improve regularization). From a clinical perspective, augmentations are also beneficial in combatting the scarcity of large clinically-acquired and annotation intensive datasets.

Due to the various benefits of modern augmentations, most importantly on the calibration of CNNs, the adoption of these techniques can be very beneficial for medical image analysis to improve the reliability and uncertainty measurements of models. Additionally, as augmentations are already used for medical image analysis and present various other advantages, modern augmentation techniques present a low barrier of entry. In comparison to other calibration techniques, modern augmentations do not effect the structure of CNN architectures and can effect calibration solely based on modifications made to datasets. Additionally, augmentations have a very low computational cost. By using modern image augmentations, major modifications do not have to be made to medical image analysis pipelines. In this study, we evaluate the performance of modern image augmentations for medical confidence calibration using various open-source medical image datasets.  

Our main \textbf{contributions} are summarized below:
\begin{enumerate}
    \item We evaluate the effects of modern augmentations on the \textit{\textbf{performance}} of CNNs for medical image analysis
    \item We understand the effects of modern augmentations on \textit{\textbf{medical confidence calibration}}
    \item We conduct experiments across various \textit{\textbf{medical modalities}} to more deeply understand the effects of modern augmentations across an array of diseases and image types
\end{enumerate}

\section{Related Work}

As follows is a review of prior work regarding standard calibration methods in the \textit{general} computer vision domain, the use of \textit{modern image augmentations} to improve CNN calibration, and the \textit{current applications} of confidence calibration methods to medical image analysis tasks.

\noindent\paragraph{Confidence Calibration:}

In a study by Gou \textit{et al.} \cite{guo2017calibration}, various observations are made on the calibration of CNNs and factors which influence this. Based on in-depth empirical experimentation, the following objective observations have been made relating to CNNs: 1. Increasing the network depth and width of CNNs typically increases accuracy \cite{zhang2021understanding} however this has negative effects on calibration. 2. Batch normalization, used for neural network optimization, often leads to miscalibration. 3. Weight decay, a regularization mechanism \cite{vapnik1999overview} commonly replaced for batch normalization, has positive effects on calibration. Popular methodologies which provide benefits in confidence calibration and quantifying predictive uncertainty include temperature scaling \cite{guo2017calibration}, bayesian neural networks \cite{denker1990transforming, mackay1992practical}, dropout as bayesian approximation \cite{gal2016dropout,srivastava2014dropout}, and ensembles of networks \cite{lakshminarayanan2017simple}. Calibration methods are evaluated based on \textit{calibration metrics} \cite{naeini2015obtaining, guo2017calibration} and \textit{reliability plotting} \cite{degroot1983comparison,niculescu2005predicting,guo2017calibration}. These two techniques provide both a quantitative and qualitative assessment of confidence calibration. 

\noindent\paragraph{Modern Augmentations for Calibration:}

The effects of modern augmentations on the regularization of CNNs is evident from the original studies of MixUp \cite{zhang2017mixup}, CutMix \cite{yun2019cutmix}, and CutOut \cite{devries2017improved}. MixUp \cite{zhang2017mixup}, an augmentation built around convex combinations of image and label pairs, was the first modern augmentation to be thoroughly studied for confidence calibration. MixUp, when benchmarked using calibration metrics, presented CNNs with significant calibration benefits according to various studies \cite{thulasidasan2019mixup, zhang2022and, carratino2020mixup}. These studies showed the benefits of MixUp, which was first proposed for regularization, to confidence calibration of CNNs. Subsequent studies covered the calibration of CNNs for the CutMix and CutOut modern augmentations \cite{chun2020empirical, das2020empirical}. These studies also concluded that these modern augmentation-based regularization techniques present CNNs with significant calibration benefits.

\noindent\paragraph{Medical Image Analysis Calibration:}
We have not identified prior studies which validate the efficacy of the MixUp, CutMix, and CutOut modern augmentations for the \textit{confidence calibration} of CNNs for medical image analysis tasks. A study by Galdran \textit{et al.} \cite{galdran2021balanced} performed experimental validation of MixUp for medical image classification. This study solely used performance-based metrics for evaluation and not calibration-based metrics. MixUp has additionally been used for medical image segmentation however, it has not been benchmarked using calibration metrics \cite{eaton2018improving}. Confidence calibration using methods other than augmentation have been studied in the medical domain \cite{liang2020improved}. In our study, we focus on understanding the effects of modern augmentations on the calibration of CNNs for medical image analysis to improve to reliability of models. Apart from augmentation-based calibration, other methods have been studied for medical image analysis calibration namely in medical image segmentation \cite{mehrtash2020confidence, thiagarajan2022training, rousseau2021post, jungo2019assessing}.

\section{Methods}

In this study, we perform experiments based on training CNNs across the MixUp, CutMix, and CutOut modern augmentations for various medical image modalities. With this, we evaluate the calibration of each CNN variant against the baseline using conventional calibration metrics and reliability plotting. The goal is to understand the effects of modern image augmentations on the confidence calibration of these CNNs. We additionally evaluate the accuracy of models using standard performance-based metrics. The formulation of the modern augmentations are briefly reviewed and described in \ref{sec:augmentations}. The metrics used to evaluate the calibration of the models are documented in Sec \ref{sec:metrics}. The various medical image modalities used are reviewed in Sec \ref{sec:datasets}. Sec \ref{sec:arch} and \ref{sec:train} review model architectures and our implementation.

\subsection{Modern Data Augmentations}\label{sec:augmentations}

\noindent\paragraph{MixUp}

Zhang \textit{et al.} \cite{zhang2017mixup} proposed MixUp as a modern augmentation technique for training neural networks on a \textit{blend} between a pair of images and labels based on convex combinations. MixUp has proven various benefits in terms of increasing robustness of neural networks when learning from corrupt labels and adversarial examples. MixUp is based on the Vicinal Risk Minimization (VRM) \cite{chapelle2000vicinal} principle, where the \textit{vicinity} of the training data distribution can be used to draw \textit{virtual} samples and shows improvements over Empirical Risk Minimization (ERM). The original formulation of MixUp from the original paper \cite{zhang2017mixup} is:

\begin{equation}
\begin{array}{l}
\tilde{x}=\lambda x_{i}+(1-\lambda) x_{j} \\
\tilde{y}=\lambda y_{i}+(1-\lambda) y_{j},
\end{array}
\end{equation}
where $\boldsymbol{x}_{i,}, \boldsymbol{y}_i$ are raw randomly sampled input vectors and $\boldsymbol{x}_{j}, \boldsymbol{y}_j$ are the corresponding one-hot label encodings. $\lambda$ are values in the range [0, 1] which are randomly sampled from the Beta distribution for each augmented example. Samples of the MixUp augmentation technique applied to various medical images are shown in Figure \ref{fig:augmentations}. 


\noindent\paragraph{CutMix}

Yun \textit{et al.} \cite{yun2019cutmix} introduced CutMix, an augmentation built upon the original formulation of MixUp and idea of combining samples. CutMix removes a patch from an image and swaps it for a region of another image generating a locally natural unseen sample. Similar to MixUp, CutMix not only combines two samples but also their corresponding labels. The formulation for CutMix is as follows:

\begin{equation}
\begin{array}{c}
\tilde{x}=M x_{i}+(1-M) x_{j} \\
\tilde{y}=\mu y_{i}+(1-\mu) y_{j},
\end{array}
\end{equation}
where $M$ indicates the binary mask used to perform the cutout and fill-in operation from two randomly drawn images. $\mu$ are values (in [0,1]) randomly drawn from the Beta distribution. Samples of the CutMix technique applied to various medical images are shown in Figure \ref{fig:augmentations}.


\noindent\paragraph{CutOut} 

This technique was proposed by DeVries \textit{et al.} \cite{devries2017improved} and is a simple augmentation technique for improving the regularization of CNNs. CutOut was formulated based on the idea of extending dropout \cite{hinton2012improving} to a spatial prior in the input space. CutOut performs occlusions of an input image similar to the idea proposed in \cite{bengio2011deep}. Rather than partially occluding portions of an image \cite{bengio2011deep}, CutOut performs fixed-size zero-masking to fully obstruct a random location of an image. CutOut differentiates from dropout as it is an augmentation technique and visual features are dropped at the input stage of the CNN whereas in dropout, this occurs in intermediate layers. The goal of CutOut is to not only improve regularization of CNNs but improve robustness to occluded samples in real-world applications. Samples of CutOut applied to medical images are shown in Figure \ref{fig:augmentations}.

\subsection{Calibration and Performance Metrics}\label{sec:metrics}

As follows are descriptions of the two techniques used to evaluate the effects of the modern augmentations on the calibration of CNNs. The first technique is a quantitative metric based on error and the second technique allows for visualizing calibration through reliability plotting.

\noindent\paragraph{Expected Calibration Error (ECE)} \cite{naeini2015obtaining} is a very widely leveraged metric for quantifying the calibration of neural networks. This approach provides a scalar summary statistic of calibration by grouping a models predictions into equally-spaced bins (\textit{B}). The weighted average of the difference between accuracy and confidence across the bins is outputted. The formulation of ECE from \cite{guo2017calibration} is as follows:

\begin{equation}
\mathrm{ECE}=\sum_{b=1}^{B} \frac{n_{b}}{N}|\operatorname{acc}(b)-\operatorname{conf}(b)|,
\end{equation}
where \textit{n} represents the number of samples. Gaps in calibration or miscalibration is represented by the difference between $\operatorname{acc}$ and $\operatorname{conf}$. In terms of the subsequently described reliability plotting, this represents the visual gaps between the identify function and plotted model calibration line. 

\noindent\paragraph{Reliability Plotting} allows for visualizing the calibration of neural networks in a qualitative manner \cite{degroot1983comparison, niculescu2005predicting}. The plot shows expected accuracy as a function of the confidence. In the case of a perfectly calibrated model, the plotted line will be identical to the identity function. Deviations from the diagonal identity function line represents miscalibrations which have occurred. The reliability diagram implementation and formulation is based on this paper \cite{guo2017calibration}.   

\noindent\paragraph{Accuracy and AUROC} are the two statistical metrics used to assess the general performance of the CNNs. Accuracy measures the fraction of predictions from the validation dataset which the model predicted correctly after training is completed. The area under the receiver operating characteristic (AUROC) is a robust measure of the ability for the binary classifier to discriminate between class labels \cite{hanley1982meaning}.

\subsection{Medical Image Datasets}\label{sec:datasets}

As follows are brief descriptions of the open-source medical image datasets used in our experiments. For training of the CNNs, 80\% of the dataset is partitioned and 20\% is used to perform the validation respectively.

\noindent\paragraph{Skin Cancer Dataset}

The Skin Cancer Dataset was sourced from the International Skin Imaging Collaboration (ISIC) organization \cite{isicarchive}. The dataset is open-source and consists of 3,297 processed dermatological skin images of mole lesions partitioned into malignant (diseased) and benign (normal) classes. Factors which differentiate images are mainly based on the pigmented skin lesions \cite{jerant2000early}.

\vspace{-12pt}

\noindent\paragraph{CXR Pneumonia Dataset}

The chest radiograph (CXR) dataset was gathered from the open-source "Chest X-Ray Images for Classification" repository from UCSD \cite{kermany2018labeled}. The dataset consists of 5,863 x-ray images (both anterior and posterior) from the normal and pneumonia classes. Images between class labels are differentiated based on hazy shadowing and opacity's found in x-rays with pneumonia.

\vspace{-12pt}
\noindent\paragraph{MRI Tumor Dataset}

The magnetic resonance imaging (MRI) dataset of the human brain was sourced from an open-source repository on Kaggle \cite{sartaj_2020}. The dataset contains 3,264 images split into tumorous and no tumor classes. The main differentiating factor between classes are the circular tumorous lesions which are typically in a difference shade compared to other regions \cite{bhattacharyya2011brain}.
\vspace{-12pt}
\noindent\paragraph{CT COVID-19 Dataset}

The CT (computed tomography) dataset of COVID-19 is from the open-source UCSD COVID-CT repository \cite{yang2020covid}. The dataset consists of 812 CT scans split into the COVID-19 positive and negative classes. COVID-19 is identified in a CT based on ground-glass opacity, vascular enlargements, and white or hazy shadowing of the lung \cite{he2020sample}.

\subsection{Model Architectures and Augmentations}\label{sec:arch}

To perform the experiments, the widely leveraged CNN architecture, ResNet is utilized \cite{he2016deep}. ResNet is applied to various medical image analysis tasks for transfer learning thus providing a robust ground-truth for experimentation \cite{kora2021transfer}. Taking into account varying CNN sizes, both ResNet-50 and ResNet-101 are benchmarked across the modern augmentation techniques. Implementations of ResNet follow the standard Keras Tensorflow \cite{tensorflow2015-whitepaper} applications plugin \footnote{\url{https://keras.io/api/applications}}. The implementation of CutMix \cite{yun2019cutmix}, CutOut \cite{devries2017improved}, and MixUp \cite{zhang2017mixup} augmentations follow open-source developments based on the original formulations \footnote{\url{https://github.com/ayulockin/DataAugmentationTF}}.

\subsection{Training Details}\label{sec:train}

All models across each dataset are trained for 100 epochs with cross entropy loss. Each dataset used contains two distinctive class labels. Thus, models are trained with two output logits for each input. Experiments are carried out using the stochastic gradient descent (SGD) optimizer \cite{kingma2014adam}, a batch size of 64, and learning rate 0.001. Input images are scaled to 224x224 pixels. For CutOut, a mask size of 50x50 pixels is used. Other augmentations follow the same parameters from open-source implementations.

\section{Results}\label{sec:results}

As follows is a summary of our systematic experimentation of modern augmentations on medical image analysis using performance-based and calibration-based evaluations across each medical image modality. Performance metrics are reported in Table \ref{table:performance} and calibration error is reported in Table \ref{table:calibration} with reliability plots displayed in Figure \ref{figure:reliability}.

\begin{table*}
\subcaptionbox{Skin Cancer}{
\begin{tabular}{|l|l|l|l|}
\hline
Model & Augmentation & Accuracy & AUROC \\ \hline
ResNet-50 \cite{he2016deep} & None & 0.792 & 0.889 \\ \hline
ResNet-50 \cite{he2016deep} & MixUp \cite{zhang2017mixup} & \textbf{0.803} & 0.890 \\ \hline
ResNet-50 \cite{he2016deep} & CutMix \cite{yun2019cutmix} & 0.797 & \textbf{0.898} \\ \hline
ResNet-50 \cite{he2016deep} & CutOut \cite{devries2017improved} & 0.801 & 0.894 \\ \hline
ResNet-101 \cite{he2016deep} & None & 0.798 & 0.885 \\ \hline
ResNet-101 \cite{he2016deep} & MixUp \cite{zhang2017mixup} & \textbf{0.825} & \textbf{0.897} \\ \hline
ResNet-101 \cite{he2016deep} & CutMix \cite{yun2019cutmix} & 0.796 & 0.889 \\ \hline
ResNet-101 \cite{he2016deep} & CutOut \cite{devries2017improved} & 0.760 & 0.879 \\ \hline
\end{tabular}
}
\hfill
\subcaptionbox{CXR Pneumonia}{
\begin{tabular}{|l|l|l|l|}
\hline
Model & Augmentation & Accuracy & AUROC \\ \hline
ResNet-50 \cite{he2016deep} & None & 0.927 & 0.944 \\ \hline
ResNet-50 \cite{he2016deep} & MixUp \cite{zhang2017mixup} & \textbf{0.944} & \textbf{0.980} \\ \hline
ResNet-50 \cite{he2016deep} & CutMix \cite{yun2019cutmix} & 0.941 & 0.977 \\ \hline
ResNet-50 \cite{he2016deep} & CutOut \cite{devries2017improved} & 0.917 & 0.941 \\ \hline
ResNet-101 \cite{he2016deep} & None & 0.872 & 0.902 \\ \hline
ResNet-101 \cite{he2016deep} & MixUp \cite{zhang2017mixup} & \textbf{0.939} & 0.976 \\ \hline
ResNet-101 \cite{he2016deep} & CutMix \cite{yun2019cutmix} & 0.933 & \textbf{0.977} \\ \hline
ResNet-101 \cite{he2016deep} & CutOut \cite{devries2017improved} & 0.886 & 0.915 \\ \hline
\end{tabular}
}
\hfill
\subcaptionbox{MRI Tumor}{
\begin{tabular}{|l|l|l|l|}
\hline
Model & Augmentation & Accuracy & AUROC \\ \hline
ResNet-50 \cite{he2016deep} & None & 0.647 & 0.684 \\ \hline
ResNet-50 \cite{he2016deep} & MixUp \cite{zhang2017mixup} & 0.607 & 0.671 \\ \hline
ResNet-50 \cite{he2016deep} & CutMix \cite{yun2019cutmix} & \textbf{0.725} & \textbf{0.825} \\ \hline
ResNet-50 \cite{he2016deep} & CutOut \cite{devries2017improved} & 0.705 & 0.748 \\ \hline
ResNet-101 \cite{he2016deep} & None & \textbf{0.705} & \textbf{0.791} \\ \hline
ResNet-101 \cite{he2016deep} & MixUp \cite{zhang2017mixup} & 0.627 & 0.738 \\ \hline
ResNet-101 \cite{he2016deep} & CutMix \cite{yun2019cutmix} & 0.607 & 0.644 \\ \hline
ResNet-101 \cite{he2016deep} & CutOut \cite{devries2017improved} & 0.568 & 0.633 \\ \hline
\end{tabular}
}
\hfill
\hfill
\subcaptionbox{CT COVID-19}{
\begin{tabular}{|l|l|l|l|}
\hline
Model & Augmentation & Accuracy & AUROC \\ \hline
ResNet-50 \cite{he2016deep} & None & \textbf{0.700} & 0.724 \\ \hline
ResNet-50 \cite{he2016deep} & MixUp \cite{zhang2017mixup} & 0.680 & \textbf{0.757} \\ \hline
ResNet-50 \cite{he2016deep} & CutMix \cite{yun2019cutmix} & 0.653 & 0.754 \\ \hline
ResNet-50 \cite{he2016deep} & CutOut \cite{devries2017improved} & 0.633 & 0.656 \\ \hline
ResNet-101 \cite{he2016deep} & None & 0.653 & 0.706 \\ \hline
ResNet-101 \cite{he2016deep} & MixUp \cite{zhang2017mixup} & \textbf{0.706} & \textbf{0.765} \\ \hline
ResNet-101 \cite{he2016deep} & CutMix \cite{yun2019cutmix} & 0.613 & 0.708 \\ \hline
ResNet-101 \cite{he2016deep} & CutOut \cite{devries2017improved} & 0.673 & 0.746 \\ \hline
\end{tabular}
}
\caption{\textbf{Performance-based metrics} of various state-of-the-art CNN models across each medical modality.}
\label{table:performance}
\end{table*}

\subsection{Skin Cancer Dataset}

\subsubsection{Performance}

The performance metrics (accuracy and AUROC) for both ResNet-50 \cite{he2016deep} and ResNet-101 \cite{he2016deep} for each augmentation technique on the skin cancer modality \cite{isicarchive} are shown in \ref{table:performance}a. For the ResNet-50 baseline on the skin cancer dataset, an accuracy of 79.2\% and AUROC of 88.9\% was achieved (Row 1). All augmentations presented minor increases in accuracy and AUROC, the most significant being MixUp \cite{zhang2017mixup} at an accuracy of 80.3\% (+1.1\% over ResNet-50). In terms of AUROC, CutMix \cite{yun2019cutmix} achieved an AUROC of 89.8\% (+0.9\% over ResNet-50). In summary, for ResNet-50, no highly significant benefits in terms of performance-based metrics were observed using modern augmentations.

For ResNet-101 on the skin cancer modality, the baseline achieved an accuracy of 79.8\% and AUROC of 88.5\%. For this model, the CutMix \cite{yun2019cutmix} and CutOut \cite{devries2017improved} augmentations performed slightly worse than the baseline in terms of accuracy. CutOut \cite{devries2017improved} also performed worse than the baseline for AUROC. MixUp \cite{zhang2017mixup} performed the best for both performance-based metrics at an accuracy of 82.5\% (+2.7\% over ResNet-101) and AUROC of 89.7\% (+1.2\% over ResNet-101). In summary, for ResNet-101, MixUp \cite{zhang2017mixup} presented fairly significant benefits in terms of performance-based metrics.

\subsubsection{Confidence Calibration}

\begin{figure*}
\centering
\vspace{-3mm}
\begin{subfigure}{\textwidth}
    \centering
    \includegraphics[scale=0.42]{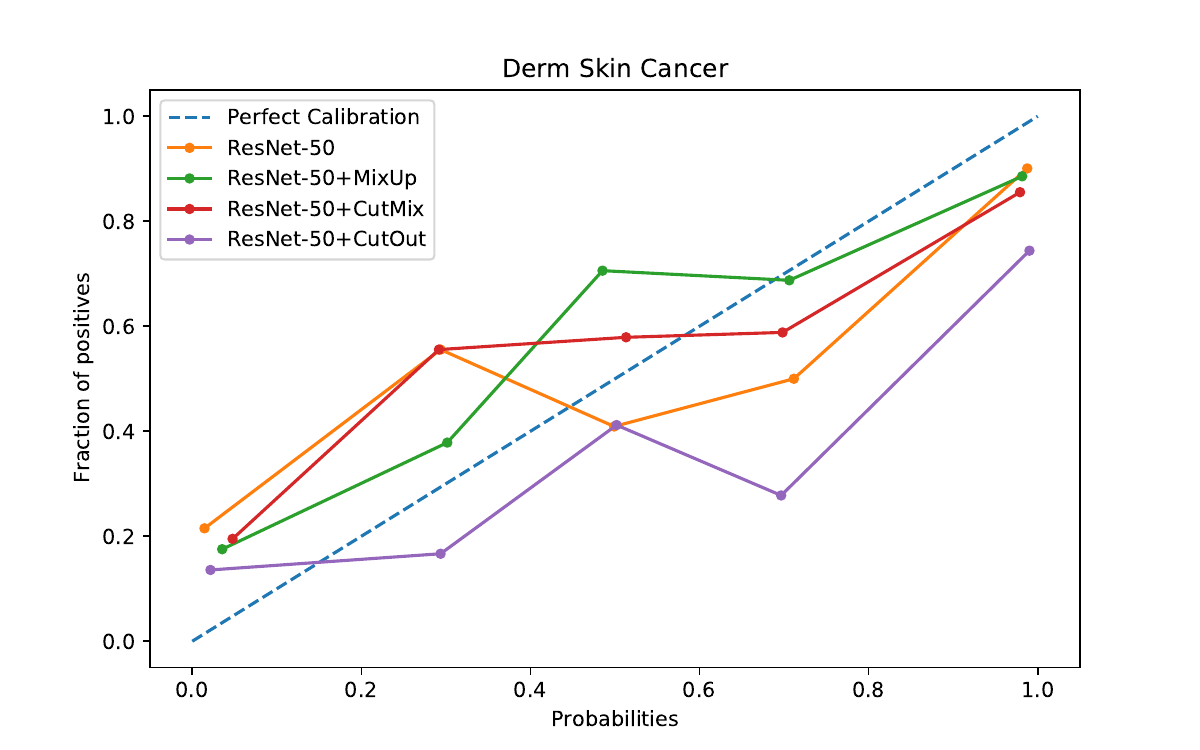}
    \includegraphics[scale=0.42]{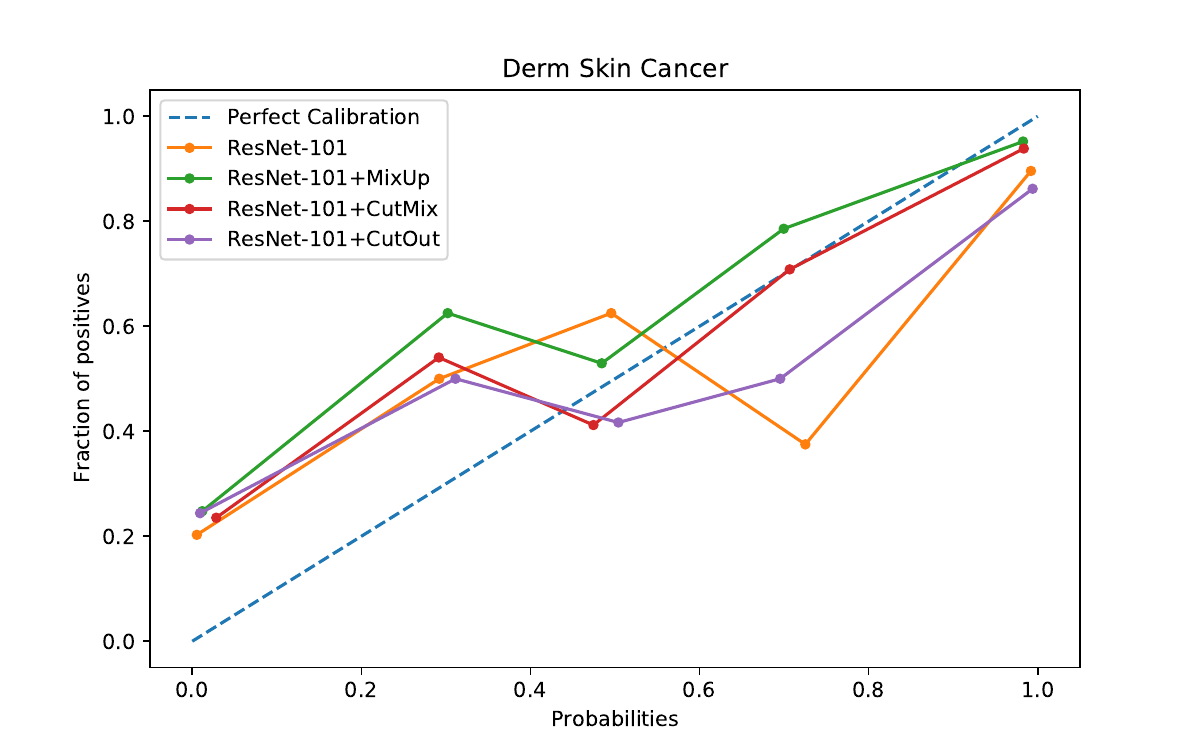}
    \vspace{-3mm}\caption{Skin Cancer}\vspace{-1mm}
    \label{SUBFIGURE LABEL 1} 
\end{subfigure}
\begin{subfigure}{\textwidth}
    \centering
    \includegraphics[scale=0.42]{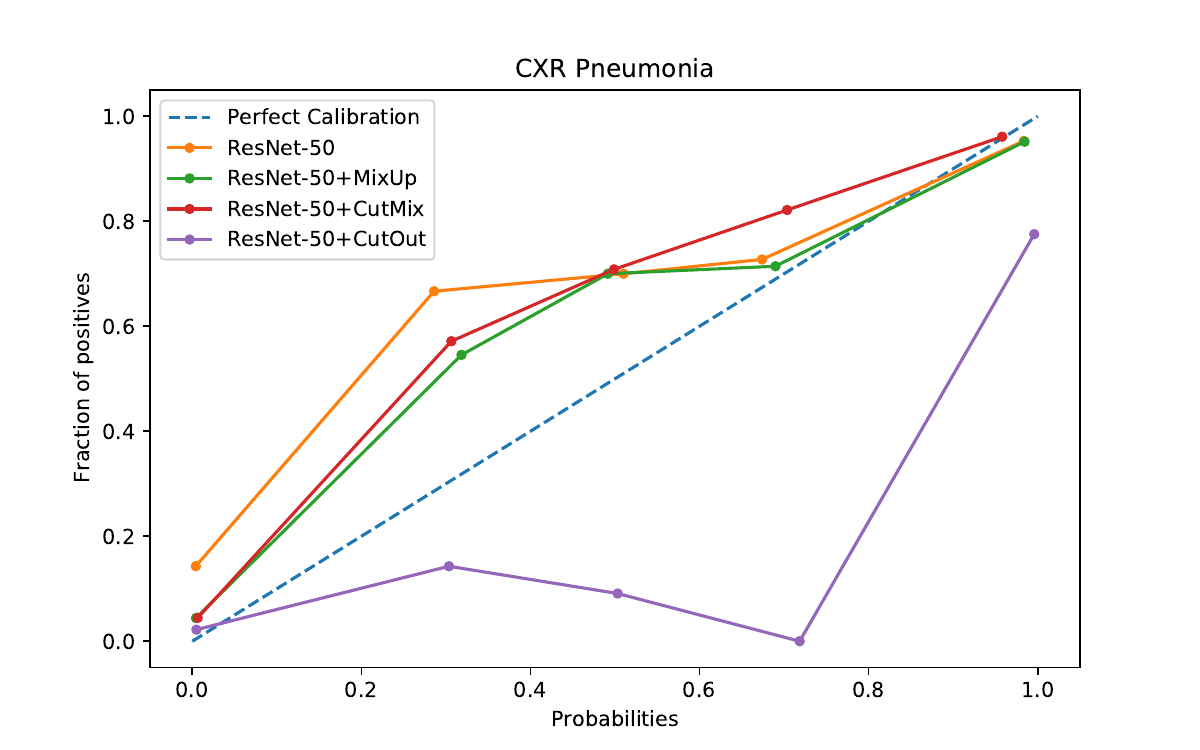}
    \includegraphics[scale=0.42]{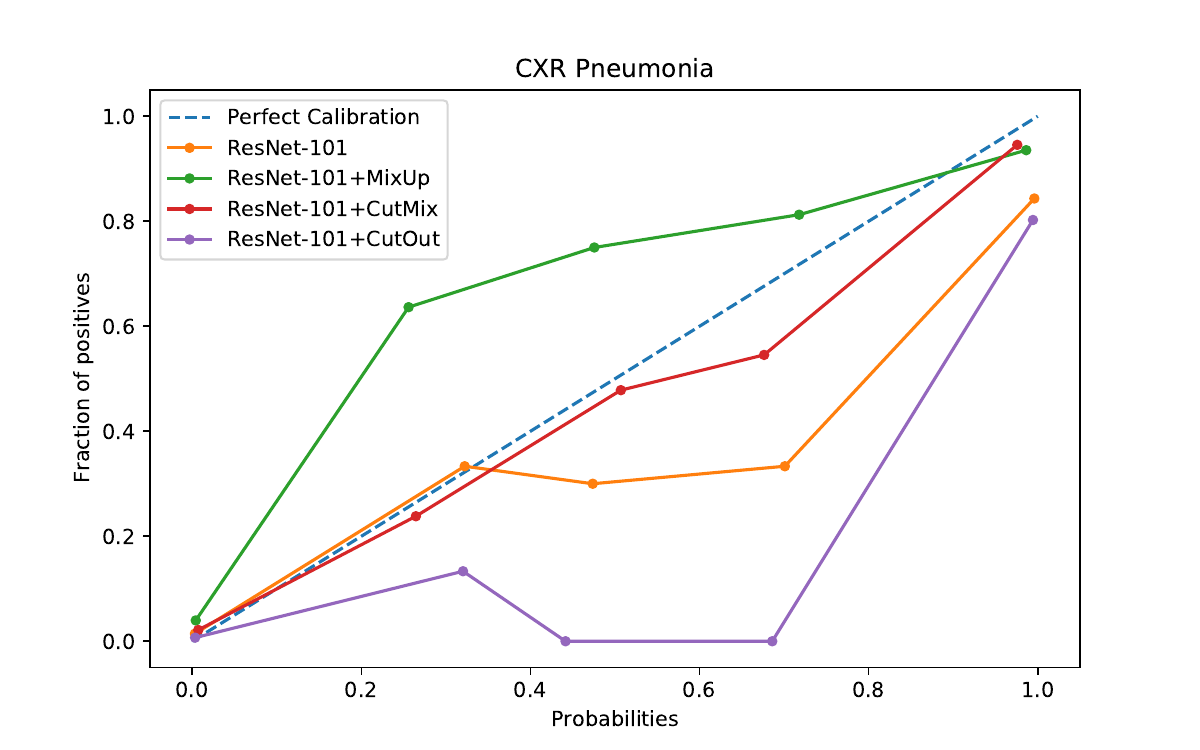}
    \vspace{-3mm}\caption{CXR Pneumonia}\vspace{-1mm}
    \label{SUBFIGURE LABEL 2}
\end{subfigure}
\\
\begin{subfigure}{\textwidth}
    \centering
    \includegraphics[scale=0.42]{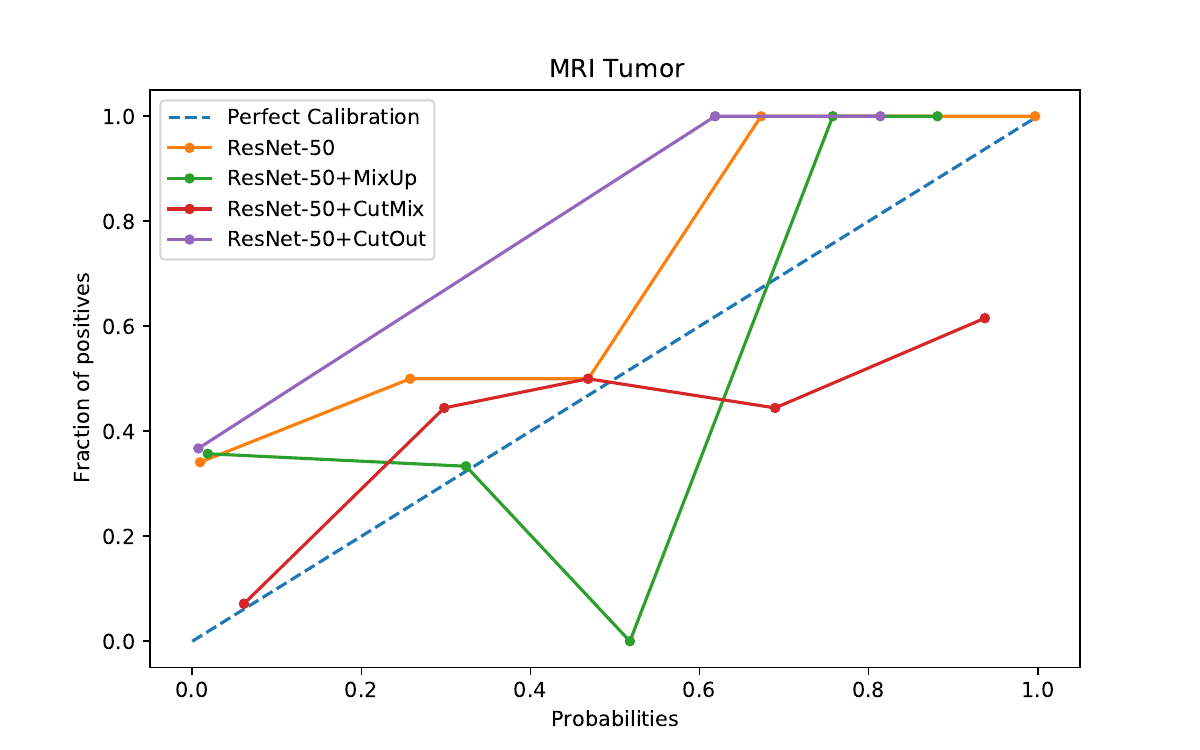}
    \includegraphics[scale=0.42]{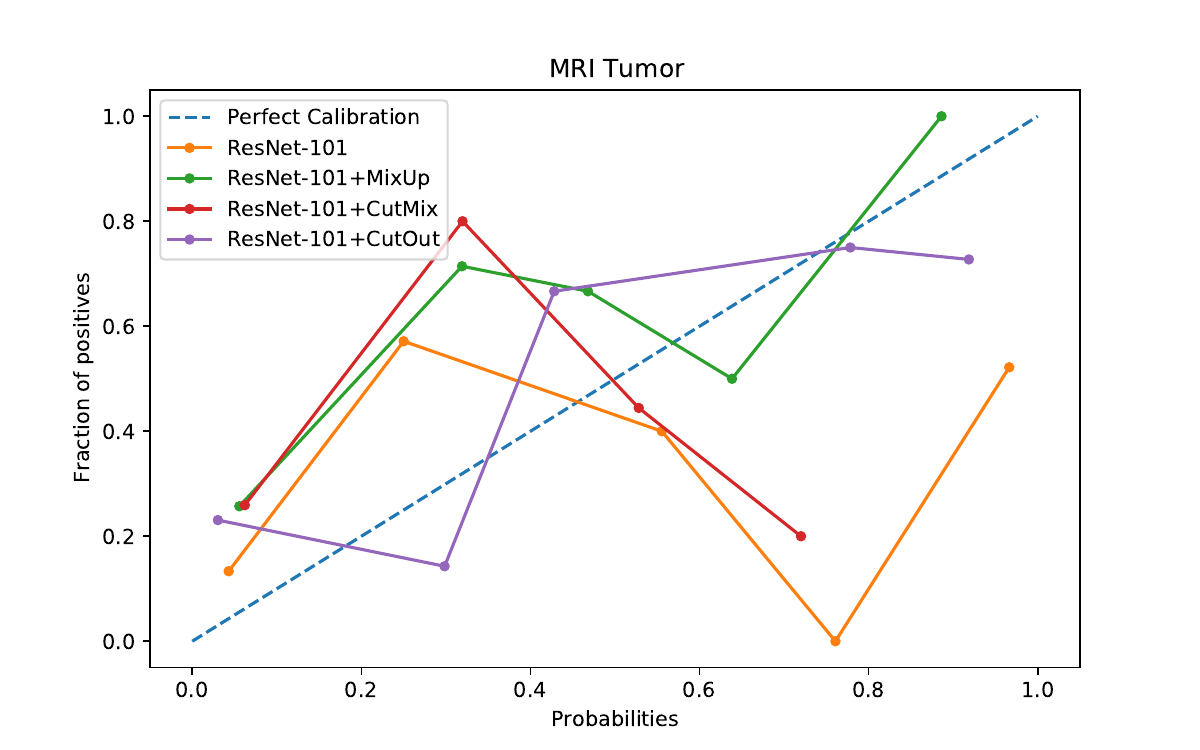}
    \vspace{-3mm}\caption{MRI Tumor}\vspace{-1mm}
    \label{SUBFIGURE LABEL 3}
\end{subfigure}
\begin{subfigure}{\textwidth}
    \centering
    \includegraphics[scale=0.42]{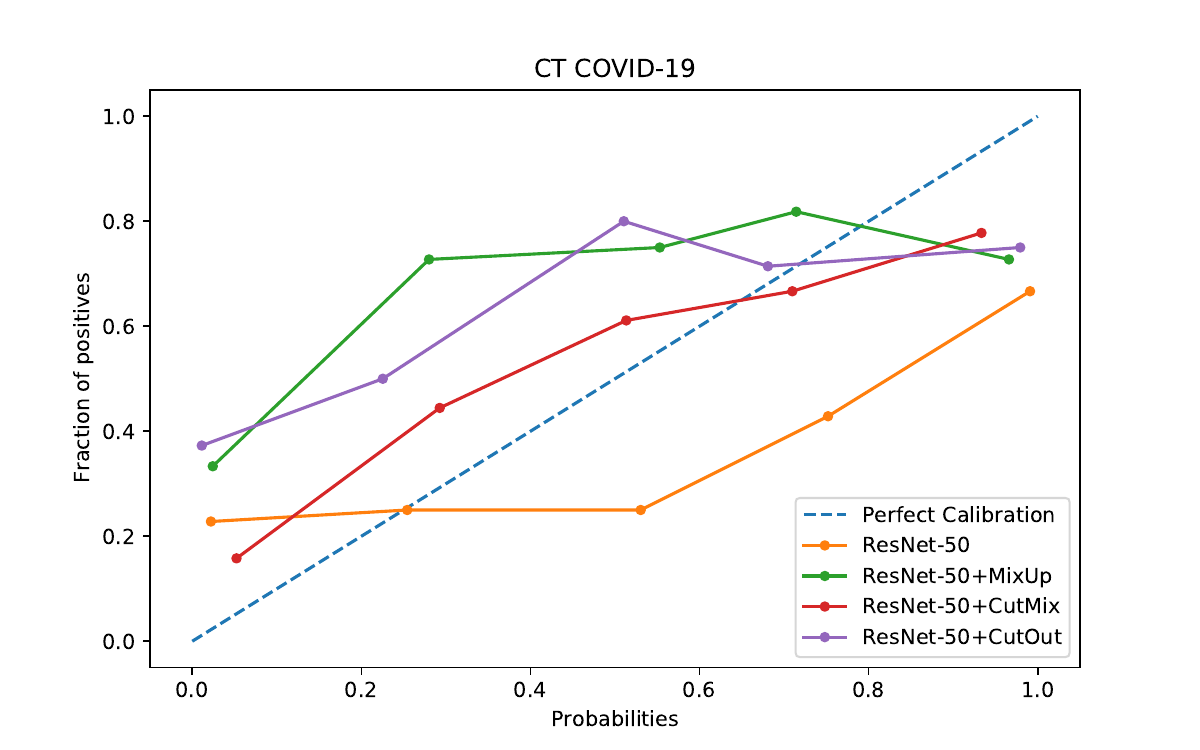}
    \includegraphics[scale=0.42]{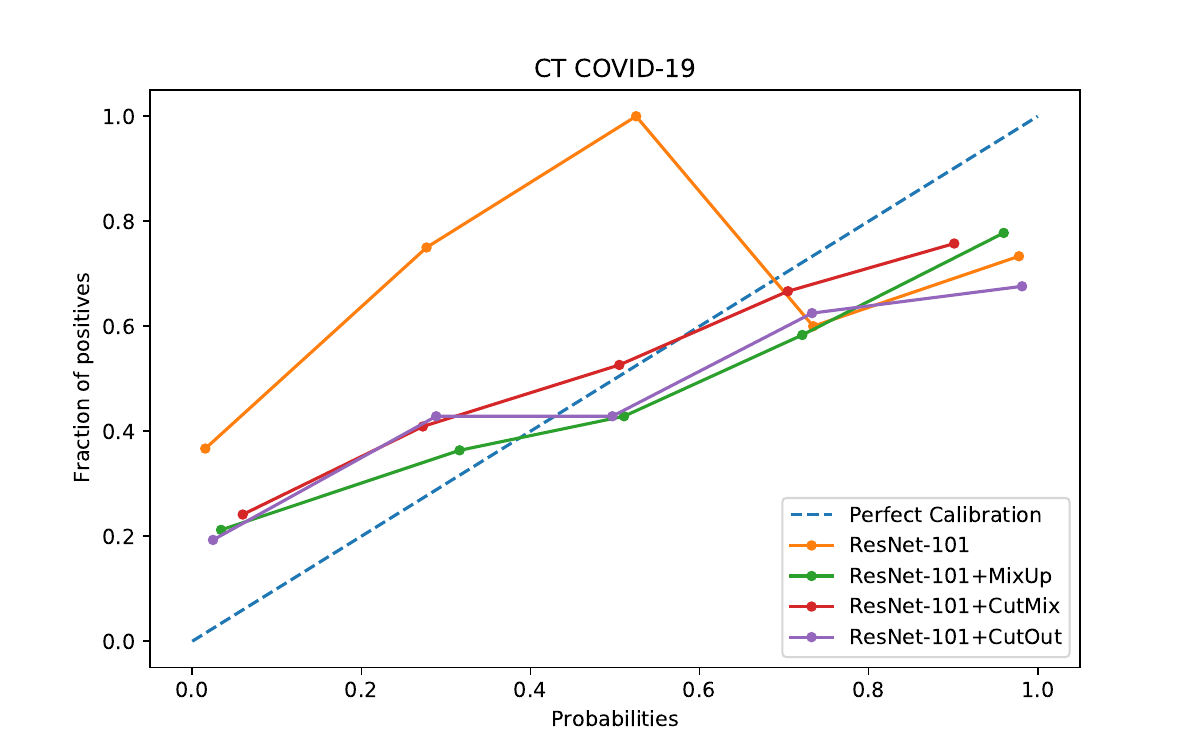}
    \vspace{-3mm}\caption{CT COVID-19}\vspace{-1mm}
    \label{SUBFIGURE LABEL 4}
\end{subfigure}
\caption{\textbf{Calibration Reliability Plots}. Shown for both models across each augmentation technique for all medical modalities. Left: ResNet-50 plot, Right: ResNet-101 plot.}
\label{figure:reliability}
\end{figure*}

The ECE calibration \cite{guo2017calibration} results for the modern augmentations on the skin cancer modality are shown in Table 2 (Row 1 and Row 2). The lower the ECE, the higher level of calibration for the model. For ResNet-50, the baseline ECE was 0.1812. All modern augmentation techniques lowered the ECE, the highest decrease was observed in CutMix \cite{yun2019cutmix} at 0.1286 (-0.0526). For ResNet-101, the baseline ECE was 0.1676. MixUp \cite{zhang2017mixup} and CutMix \cite{yun2019cutmix} both lowered the ECE however, CutOut \cite{devries2017improved} increased the ECE to 0.1967 (+0.0291). The most significant decrease in ECE for ResNet-101 was observed in CutMix \cite{yun2019cutmix} at 0.0973 (-0.0703). For both models, across the augmentations, CutMix \cite{yun2019cutmix} presented the most significant decreases in ECE providing higher calibration. The reliability plots \cite{guo2017calibration} for the skin cancer modality are shown in Figure \ref{figure:reliability}a.

\begin{table*}
\centering
\begin{tabular}{|l|l|l|l|l|l|}
\hline
Dataset & Model & Baseline & MixUp \cite{zhang2017mixup} & CutMix \cite{yun2019cutmix} & CutOut \cite{devries2017improved} \\ \hline
Derm & ResNet-50 \cite{he2016deep} & 0.1812 & 0.1424 (-0.0388) & \textbf{0.1286 (-0.0526)} & 0.1726 (-0.0086) \\ \hline
Derm & ResNet-101 \cite{he2016deep} & 0.1676  & 0.1020 (-0.0656)  & \textbf{0.0973 (-0.0703)}  & 0.1967 (+0.0291)  \\ \hline
CXR & ResNet-50 \cite{he2016deep} & 0.0675  & 0.0409 (-0.0266) & \textbf{0.0351 (-0.0324)} & 0.0750 (+0.0075) \\ \hline
CXR & ResNet-101 \cite{he2016deep} & 0.1150 & \textbf{0.0340 (-0.081)} & 0.0448 (-0.0702) & 0.1024 (-0.0126) \\ \hline
MRI & ResNet-50 \cite{he2016deep} & 0.3419 & 0.3675 (+0.0256) & \textbf{0.1259 (-0.2416)} & 0.2874 (-0.0801) \\ \hline
MRI & ResNet-101 \cite{he2016deep} & \textbf{0.2665} & 0.3675 (+0.101) & 0.3487 (+0.0822) & 0.3770 (+0.1105) \\ \hline
CT & ResNet-50 \cite{he2016deep} & 0.2866 & 0.2361 (-0.0505) & \textbf{0.1909 (-0.0957)} & 0.3367 (+0.0501) \\ \hline
CT & ResNet-101 \cite{he2016deep} & 0.3237 & \textbf{0.1975 (-0.1262)} & 0.2382 (-0.0855) & 0.2464 (-0.0773) \\ \hline
\end{tabular}
\caption{\textbf{Expected Calibration Error (ECE)} (\textit{M = 15 bins}) across various medical imaging modalities and CNN architectures.}
\label{table:calibration}
\end{table*}

\subsection{CXR Pneumonia Dataset}
\subsubsection{Performance}

The performance-based metrics for the CXR modality \cite{kermany2018labeled} are reported in Table \ref{table:performance}b. The ResNet-50 baseline performed with an accuracy of 92.7\% and AUROC of 94.4\%. Both MixUp \cite{zhang2017mixup} and CutMix \cite{yun2019cutmix} presented significant increases in both accuracy and AUROC however, CutOut \cite{devries2017improved} decreased both accuracy and AUROC. The most significant increase in the performance-based metrics was observed in MixUp \cite{zhang2017mixup} with an accuracy of 94.4\% (+1.7\%) and AUROC of 98\% (+3.6\%). In summary, MixUp \cite{zhang2017mixup} presented the highest benefits in terms of performance.

For ResNet-101, the baseline accuracy was 87.2\% and AUROC was 90.2\%. All augmentations presented increases in both performance-based metrics. The highest increase in accuracy was observed in MixUp \cite{zhang2017mixup} at 93.9\% (+6.7\%). The highest increase in AUROC was observed in CutMix \cite{yun2019cutmix} at 97.7\% (+7.5\%). In summary, both MixUp \cite{zhang2017mixup} and CutMix \cite{yun2019cutmix} presented increases in performance.

\subsubsection{Confidence Calibration}

The ECE calibration results for the modern augmentations on the CXR pneumonia modality are shown in Table \ref{table:calibration} (Row 2 and Row 3). The ResNet-50 baseline had an ECE of 0.0675. Both MixUp \cite{zhang2017mixup} and CutMix \cite{yun2019cutmix} decreased the ECE. However, CutOut \cite{devries2017improved} increased the ECE slightly. The lowest ECE was observed for CutMix at 0.0351 (-0.0324). ResNet-101 had a baseline ECE of 0.1150. All augmentations reduced the ECE, MixUp \cite{zhang2017mixup} had the most significant decrease at 0.0340 (-0.081). In summary, both MixUp \cite{zhang2017mixup} and CutMix \cite{yun2019cutmix} presented the most benefits for calibration. The reliability plots displaying the level of calibration for the CXR pneumonia modality are shown in Figure \ref{figure:reliability}b for both the pneumonia and normal class labels.

\subsection{MRI Tumor Dataset}
\subsubsection{Performance}
The performance-based metrics for the MRI modality are reported in Table \ref{table:performance}c. For ResNet-50, the baseline performed with an accuracy of 64.7\% and AUROC of 68.4\%. In this scenario, MixUp \cite{zhang2017mixup} reduced the performance for both accuracy and AUROC while CutMix \cite{yun2019cutmix} and CutOut \cite{devries2017improved} increased the performance. The highest performing augmentation was observed in CutMix \cite{yun2019cutmix} at an accuracy of 72.5\% (+7.8\%) and AUROC of 82.5\% (+14.1\%).

For ResNet-101, the baseline performed with an accuracy of 70.5\% and AUROC of 79.1\%. All calibration metrics performed with lower performance-based metrics leaving the baseline as the highest performing model. The highest decrease in performance was observed in CutOut \cite{devries2017improved} at an accuracy of 58.6\% (-11.9\%) and AUROC of 63.3\% (-15.8\%).

\subsubsection{Confidence Calibration}

The ECE calibration results for the modern augmentations on the MRI tumor modality are shown in Table \ref{table:calibration} (Row 5 and Row 6). For ResNet-50, the baseline ECE was 0.3419. In this scenario, CutMix \cite{yun2019cutmix} and CutOut \cite{devries2017improved} lowered the ECE and MixUp \cite{zhang2017mixup} increased the ECE to 0.3675 (+0.0256). The highest decrease in ECE was observed for CutMix \cite{yun2019cutmix} at an ECE of 0.1259 (-0.2416). ResNet-101 has a baseline of 0.2665. Interestingly, all augmentations increased the ECE the most significant increase was observed in CutOut \cite{devries2017improved} at 0.3770 (+0.1105). The baseline for ResNet-101 has the lowest ECE. The reliability plots for the MRI tumor modality are shown in Figure \ref{figure:reliability}c.

\subsection{CT COVID-19 Dataset}
\subsubsection{Performance}

The performance-based metrics for the CT modality \cite{yang2020covid} are reported in Table \ref{table:performance}d. For ResNet-50, the baseline performed with an accuracy of 70.0\% and AUROC of 72.4\%. In terms of accuracy, all augmentations presented lower accuracy compared to the baseline the most significant observed in CutOut \cite{devries2017improved} with an accuracy of 63.3\% (-6.7\%). For AUROC, both MixUp \cite{zhang2017mixup} and CutMix \cite{yun2019cutmix} increased performance and CutOut \cite{devries2017improved} reduced the AUROC to 65.6\% (-6.8\%). The highest increase in AUROC was observed in MixUp \cite{zhang2017mixup} at 75.7\% (+3.3\%).

For ResNet-101, the baseline performed with an accuracy of 65.3\% and AUROC of 70.6\%. In terms of accuracy, MixUp \cite{zhang2017mixup} and CutOut \cite{devries2017improved} presented increases while CutMix decreased the accuracy to 61.3\% (-4\%). The highest increase in accuracy was observed in MixUp \cite{zhang2017mixup} at an accuracy of 70.6\% (+5.3\%). For AUROC, all augmentations presented increases in performance the most significant observed in MixUp \cite{zhang2017mixup} at an AUROC of 76.5\% (+5.9\%).

\subsubsection{Confidence Calibration}

The ECE calibration results for the modern augmentations on the CT COVID-19 modality are shown in Table \ref{table:calibration} (Row 7 and Row 8). For ResNet-50, the baseline ECE was 0.2866. In this scenario, MixUp and CutMix reduced the ECE while CutOut increased the ECE (+0.0501). The most significant ECE decrease was observed in CutMix at an ECE of 0.1909 (-0.0957). ResNet-101 had an ECE baseline of 0.3237. All augmentations reduced the ECE the most significant decrease was observed in MixUp at 0.1975 (-0.1262). Reliability plots for the CT modality are shown in Figure \ref{figure:reliability}

\subsection{Interpretation}

In summary, it is evident that in certain scenarios of medical image analysis, modern image augmentations can increase performance and significantly improve confidence calibration of CNNs. It is also important to understand that certain modern augmentations can decrease the performance and lead to miscalibration of CNNs. Table \ref{table:review} shows a numerical summary of the amount of times a specific modern augmentation increased or decreased the level of calibration across all experiments. 

\begin{table}[h!]
\centering
\begin{tabular}{|l|c|c|}
\hline
Augmentation & \multicolumn{1}{l|}{↑Calib} & \multicolumn{1}{l|}{↓Calib} \\ \hline
MixUp \cite{zhang2017mixup} & 6 & 2 \\ \hline
CutMix \cite{yun2019cutmix} & 7 & 1 \\ \hline
CutOut \cite{devries2017improved} & 4 & 4 \\ \hline
\end{tabular}
\caption{Numerical summary of calibration results from experimentation for augmentations across all modalities.}
\label{table:review}
\end{table}

From these results, it is evident that CutMix increased the level of calibration (decreased expected calibration error) most frequently (7 out of 8 times). MixUp also presented significant impact on calibration having increased the level of calibration for 6 out of 8 experiments. However, CutOut increased calibration only 4 out of 8 times. Such performance shows that not all modern augmentations can positively effect calibration. CutOut could potentially be detrimental to the calibration of CNNs for medical image analysis tasks. We hypothesize that the reason for CutOut reducing performance is because it can potentially remove clinically relevant regions from the images. On the other hand, MixUp and CutMix modify visual information but do not remove regions completely.

\section{Conclusion}

In this paper, we have compared the effects of several modern data augmentations on the confidence calibration of CNNs for various medical image analysis tasks using open-source datasets. CNNs are often prone to overconfidence and unreliable uncertainty estimates leading to a low amount of awareness. Improper quantification of uncertainty can be a high risk in the clinical setting and could potentially lead to medical errors. Through our in-depth experiments on the calibration of CNNs for medical image analysis using both ECE and reliability plotting, it is evident that certain modern augmentations (e.g. MixUp and CutMix) present significant benefits in terms of calibration, while others (CutOut) could worsen performance. Additionally, through the use of conventional performance-based metrics, it is evident that modern augmentations can also significantly increase the accuracy of CNNs. In conclusion, the usage of modern augmentations in medical image analysis CNNs can be beneficial in improving the reliability of models to improve clinical decision-making, however they should be benchmarked before implementation.

{\small
\bibliographystyle{ieee_fullname}
\bibliography{egbib}
}

\end{document}